\documentstyle[12pt,aaspp4,flushrt]{article}

\begin{document}
\def\mso{\,\mathrm{M}_\odot}
\def\rso{\,{\rm R}_\odot}
\def\lso{\,{\rm L}_\odot}
\def\teff{\log\, T_{\rm eff}\,}
\def\simle{\mathrel{\hbox{\rlap{\hbox{\lower4pt\hbox{$\sim$}}}\hbox{$<$}}}}
\def\simgr{\mathrel{\hbox{\rlap{\hbox{\lower4pt\hbox{$\sim$}}}\hbox{$>$}}}}

\title{R4 and its circumstellar nebula: evidence for a binary merger?$^1$}
\footnotetext[1]{Based on observations obtained at the European 
                 Southern Observatory, La Silla}
\author{A. Pasquali}
\affil{ESO/ST-ECF, Karl-Schwarzschild-Strasse 2, D-85748 Garching bei 
M\"unchen, Germany}
\authoremail{apasqual@eso.org}
\author{A. Nota $^{2}$}
\footnotetext[2]{Affiliated with the Astrophysics  
        Division, Space Science Department of the European Space Agency}
\affil{STScI, 3700 San Martin Drive, Baltimore, MD 21218, USA}
\authoremail{nota@stsci.edu}
\author{N. Langer}
\affil{Institut f\"ur Physik, Universit\"at Potsdam, Postfach 601553,
D-14415 Potsdam, Germany}
\authoremail{ntl@leo.astro.physik.uni-potsdam.de}
\author{R.E. Schulte-Ladbeck}
\affil{University of Pittsburgh, Pittsburgh, PA 15260}
\authoremail{rsl@phyast.pitt.edu}
\author{M. Clampin}
\affil{STScI, 3700 San Martin Drive, Baltimore, MD 21218, USA}
\authoremail{clampin@stsci.edu}

\begin{abstract}
We present new, NTT longslit spectroscopy of the B[e] supergiant in the
binary system R4 in the Small Magellanic Cloud. The data show 
extended, forbidden N and S emissions which are typical signatures
of circumstellar matter. Their extension along the space axis of the
slit defines an angular size of 8.6$''$ which 
translates into a linear size of 2.4 pc. 
The N emission lines also show the velocity  
structure of a bipolar outflow expanding at 100 km s$^{-1}$
on average. This implies that, for a measured radius of 1.2 pc, the
outflow originated about 1.2 $\times$ 10$^4$ yr ago. The line flux ratio
[NII]6584/[SII]6717 indicates that the nebula is nitrogen enriched and
therefore it has been ejected from the central star. 
\par
This is the first bipolar, ejection nebula detected around a well-established
B[e] supergiant. 
The bipolar morphology and the chemical enrichment shown by the nebula
associated with R4 are consistent with the picture of a binary merger
(Langer \& Heger 1998), in which R4 was originally a system composed by a close
pair and a third star (the observed A companion). The close pair merged
into a single star and the merging process produced a circumstellar
nebula that was later shaped by the ensueing B star wind. 
\par
Since the bipolar morphology, the
kinematics and the enriched chemical composition make the nebula
surrounding R4 very similar to the observed LBV nebulae, 
our findings imply that at least a few LBV outbursts and nebulae might well be 
the result of the merging process of two massive stars.  
\end{abstract}


\section{Introduction}
The class of B[e] supergiants consists of about 20 luminous evolved B~stars
with a rich emission line spectrum and a strong infrared excess (Zickgraf
et al. 1986, Lamers et al. 1998). Most of the confirmed members of this class
are located in the Magellanic Clouds,  mainly for two reasons: the 
luminosities of the 
Galactic objects cannot be precisely determined due to the uncertain 
distances, and the 
difficulty to resolve  the objects of this class from other B-type emission 
line stars 
(Be~stars, Herbig Be~stars, and other types of B[e] stars).

Gummersbach et al. (1995) were able to place 14 Magellanic Cloud
B[e] supergiants in the HR~diagram. There, they appear to define two distinct
groups, one at relatively low luminosity ($L \simle 10^5\lso$) and low
effective temperature ($\teff \simle 15\, 000\,$K), and the other at
higher luminosities ($L \simgr 10^5\lso$) and temperatures
($\teff \simeq 12\, 000\,$K -- $30\, 000\,$K). 

The spectral properties of the B[e] supergiants are best explained within
the model by Zickgraf et al. (1985), who propose that these stars have
a two component wind: a fast polar wind responsible for the high excitation
UV resonance lines, and an equatorial slow and cool wind producing the narrow 
permitted and forbidden lines. The equatorial wind is associated with
the highest mass-loss rate and usually identified with an outflowing disk 
where dust can condense and emit at infrared wavelengths.
Such disk might be produced by rotational wind compression
(Bjorkman \& Cassinelli 1993, Bjorkman 1999).
Nevertheless, it remains to be shown that disk inhibition due to
non-radial components of the wind driving line force and 
gravity darkening (Owocki \& Gayley 1998) can be overcome, 
perhaps by a combination of rotational
compression and wind bi-stability proposed by Lamers \& Pauldrach (1991)
which predicts a sudden increase in the wind mass flux and decrease in the
wind velocity at a critical temperature ($\sim 20\, 000\,$K) 
when the stellar surface temperature decreases gradually from the pole 
towards the equator.

Langer \& Heger (1998) have connected the B[e] supergiant
stage with phases in the evolution of rotating
massive stars during which the star can possibly reach the $\Omega$-limit,
i.e. its surface rotation rate (which also takes into account the radiation 
force of the star) is able to  
destabilize the stellar surface at the equator (Langer
1997). They found that the most luminous and hot B[e] stars
might be related to core hydrogen burning models which arrive at the
$\Omega$-limit due to increasing surface opacities during their main sequence
evolution, which is possible for stars relatively close 
to the Eddington-limit even if they are slow rotators (Langer 1998).
They proposed further that stars below $\sim 10^5\lso$
could reach the $\Omega$-limit during core helium burning (on the so called
blue loops) due to efficient angular momentum transport from the stellar
interior to the stellar surface during this phase (Heger \& Langer 1998).
Finally, the outbursts of Luminous Blue Variables
have been associated with these stars hitting the $\Omega$-limit
(Langer 1997, Langer et al. 1999), a conjecture
which is strongly supported by the bi-polarity of virtually all 
circumstellar nebulae of LBVs (Nota et al. 1995).

Whether all massive stars go through a B[e] supergiant stage, and whether 
they are connected to Luminous Blue Variables is unclear. Empirically,
the distribution of the group of luminous B[e] supergiants in the
HR~diagram overlaps with that of the LBVs (Bohannan 1997).
A connection between B[e] supergiants and LBV stars has
been early suggested by Shore (1990) and Schulte-Ladbeck \& Clayton
(1993) from their analysis of S22, in the Large Magellanic Cloud.
Classified as a B[e] supergiant by Zickgraf et al. (1986), S22 shows an
intrinsic polarization of 0.52 $\%$ due to electron scattering in an
aspherical wind.
The polarization degree is
variable and this is probably linked to variations in the
mass-loss rate of the star (Schulte-Ladbeck \& Clayton 1993). A similar
result has been found for the galactic LBV HR Carinae,
which is characterized by an intrinsic continuum
polarization of about 0.4$\%$, possibly variable (Clampin et al. 1995).
This can again be
explained as due to a non-spherical wind geometry (the presence of a
circumstellar disk has been also discussed by Nota et al. 1997) and a
time dependent mass loss rate. In addition, Shore (1990) has detected
almost a  factor of two variation in the UV flux of S22 longward of 1600
\AA\ and a factor between 2 and 3 variation shortward of 1600 \AA. The
amplitude of the UV variability is quite similar to that
observed in LBVs during their shell ejection phase (Pasquali \& Nota
1999).

As an alternative approach, to study the occurrence of the LBV phase in
the evolution of massive stars,
we have undertaken a longslit spectroscopy campaign of
galactic and MC evolved supergiants whose stellar properties (M$_{Bol}$ and Log
T$_{eff}$) are in the range set by confirmed LBVs. The aim of the
observations is
to detect the presence of circumstellar nebulae and
to determine whether these are ejected by the star and possibly
establish an evolutionary connection with LBVs.

Here, we present the first results obtained for the R4,
in the Small Magellanic Cloud. 
With $L\simeq 10^5\lso$ and T$_{eff} \simle 27\, 000\,$K 
(Zickgraf et al. 1996), R4 is the hottest
and least luminous star within the high luminosity group of B[e] supergiants.
Zickgraf et al. showed that R4 is a spectroscopic
binary (a = 23 A.U.) comprising a B[e] supergiant with 
spectrophotometric variability characteristic of a LBV, and an evolved A~type 
companion star which is about 10~times less luminous ($10^4\lso$).
In Sect.~2 we present the spectroscopic data taken for R4, while in Sect.~3 
we describe the results obtained from our observations. A discussion
of our results, of  the implications for the evolutionary history
of R4, and  of the connection of B[e] supergiants with LBVs 
follows in Sect.~4.

\section{Observations and data reduction}
We observed R4 at the 3.5m ESO/NTT telescope, on the nights of July 27 -
30, 1998. We used the EMMI spectrograph in the REMD (Red Medium
Dispersion) configuration and acquired longslit spectra through gratings 
6 and 7 in the following wavelength ranges: 4470 - 5820 \AA, 5080 - 
6420 \AA\ and 6240 - 6870 \AA. We observed in three different positions:
on the star, at 3$''$ North and 3$''$ South, respectively. These two
offsets  were computed by translating the typical linear sizes of
known LBV nebulae into angular sizes at the SMC distance (57.5 Kpc, van
den Bergh 1989), in order to be
able to detect a circumstellar nebula, if present, when no
high-resolution imaging could be performed. We employed a 1$''$ x
180$''$ longslit  oriented East - West in all the cases but in the
4470 - 5820 \AA\ spectrum, for which the slit had been oriented North -
South. The complete journal of the observations is reported  in Table 1, 
where the wavelength range, the dispersion and effective
resolution, the exposure time and the offset are summarized for each spectrum.
The red arm of EMMI was equipped with a TK2048EB4 CCD, whose spatial
scale is 0.27$''$ per pixel. 
\par\noindent
We obtained the usual set of
bias and flat-field images, together with comparison spectra of HeAr for the
wavelength calibration. 
The spectra were first cleaned to remove cosmic rays and bad columns, and
then corrected for bias and flat-fielded. Their subsequent reduction was
performed following the procedure outlined in the IRAF LONGSLIT package.
We modeled the sky background in each frame by fitting a surface 
described by Chebyshev polinomials of low order in X and Y. This surface was 
then subtracted from the frame itself. The wavelength calibration 
was achieved in three steps: first, we identified the emission lines at
the central row of the comparison spectrum and derived the dispersion
correction by fitting a Chebyshev function of order 3. The same lines
were then reidentified across the entire frame adopting a step of 2 rows
and readjusting the dispersion correction if necessary. Finally, the
dispersion corrections obtained through the frame were fitted into a
surface using Chebyshev polynomials of order 5. This surface
was then applied to each spectrum for the final wavelength calibration.
The same procedure was of course repeated for each grating. The
effective resolution is reported in Column 5 of Table 1 for each
grating: a FWHM = 2.72 and 2.36 \AA\ corresponds to a velocity resolution 
of 84 and 61 km s$^{-1}$ 
at H$\beta$ and [NII]$\lambda$5755 respectively, while a FWHM of 1.12
\AA\ defines a velocity resolution of 25 km s$^{-1}$ at H$\alpha$.
Measurements of the sky lines indicate that the wavelength calibration
is certain at $\pm$ 10 km s$^{-1}$ at any position within the longslit
spectra.  
\par\noindent

\section{Results}
\subsection{Nebular kinematics}
The long-slit spectra  of the R4 region show the presence of extended
nebular lines, such as [NII] $\lambda\lambda$6548, 6584 and [SII]
$\lambda\lambda$6717, 6731.  These lines show a well defined spatial
extension and velocity structure, clearly distinguishable from  the
underlying emission due to the local interstellar medium.  We show  in
Figure 1 the [NII] $\lambda$6584 line as detected  at the position
3$''$ North  with respect to the star (top panel), and 3$''$ South
(bottom panel). The spatial extent of the  nebular emission is
approximately 32 pixels, which corresponds to an angular size of
8.6$''$ (given the EMMI/CCD spatial scale  of 1 pixel = 0.27$''$) and
to  a linear size of 2.4 pc, in the assumption of a distance for the
SMC of 57.5 Kpc  (van den Bergh 1989). The  peculiar velocity
structure  of the nebular lines can be immediately noted in a quick
look inspection of the two-dimensional spectra:  while at 3$''$ North
the bulk of the emission is redshifted,  at 3$''$ South the line is
blueshifted. A remarkable symmetry is present, both along the spatial
direction (North  compared with  South), and in the velocity  structure
(redshifted versus blueshifted).

We used the [NII]$\lambda$6584 line to derive the nebular radial
velocity map at the two offset pointings.  We binned the spectra by a
factor of 2 along the spatial axis (2 pixels corresponding to 0.54$''$)
and extracted an individual spectrum from each bin. We measured the
peak wavelengths in the [NII] profile by multi-gaussian fitting and
computed the corresponding radial expansion velocities. Our fits are
characterized by a typical error of $\pm$ 4 km s$^{-1}$.  In this
wavelength region, the spectra have a velocity resolution of $\simeq$
25 km s$^{-1}$ and the  absolute wavelength calibration is within $\pm$
10 km s$^{-1}$  across the  entire slit.  The  derived radial
velocities, corrected for the heliocentric motion, have  been  plotted
in Figure 2 as a function of distance from the star, in arcseconds, for
both slit positions (3$''$ North  - top panel;  3$''$ South -
bottom panel). On the abscissae, East is to the left and West is to the
right. The star is at position 0. 

From Figure 2 it is clear that  the local interstellar medium
dominates the velocity distribution at distances  larger than $|$5$|''$
from the star, since we measure the  mean radial velocity  of this overall
motion  to be 110 km s$^{-1}$, in agreement with the Fabry-Perot
H$\alpha$ observations of le Coarer et al. (1993). However, in the
distance range -5$'' <$ d $< 5''$ from the star,  two additional
components are  clearly resolved: at the position  3$''$ North we find  a
component  which is  redshifted, and spans between 200 and 280
km s$^{-1}$, covering the entire spatial range.  At the position 3$''$ South,
a  second component  is also present, which varies between 20
and 90  km s$^{-1}$. This component extends between 4$''$ E and 4$''$ W
from the star. 

The remarkable symmetry  of these radial velocity structures can be
better appreciated in Figure 3 (top panel) where we have plotted on the same
spatial scale the two velocity profiles obtained at the two positions.
First, we  notice that the northern velocity distribution mirrors the 
southern. 
The N component  displays two radial velocity maxima at $\simeq$ 2.5$''$ E
and $\simeq$ 1.5$''$ W. In correspondence  to the same  two positions,
the S component shows two radial velocity minima. The N component
reaches radial velocity minima in correspondence  to 
the star and at the outer boundaries of its spatial extension. The S
component follows a symmetrical trend.  The peak-to-peak  radial
velocity amplitude of both components is very similar ($\simeq$  80 km
s$^{-1}$).  In addition, there is symmetry between the E and W regions of the 
nebula
with respect to the central star. Indeed, the eastern portion of both velocity 
curves, 
when folded, significantly overlaps the western. 

The observed velocity structure is likely to be indicative of a nebula
surrounding R4. The nebula is dynamically associated with the central
star, for which  Zickgraf et al. (1996) determined a radial velocity of
147 km s$^{-1}$, and most likely has been ejected by R4 in a previous phase.
Therefore, with respect to the central star, the northern component of
the nebula turns out to be red-shifted by 84 km s$^{-1}$ while the
southern is blue-shifted by 118 km s$^{-1}$ on average. Assuming a mean
expansion velocity of 100 km s$^{-1}$ and a full linear size of 2.4 pc,
we derive a dynamical age of the nebula of $\sim$ 1.2 $\times$ 10$^4$ yr.
\par\noindent
Without a direct image, and on the basis of kinematics considerations alone, 
it is difficult to make definite conclusions on its structure.
Compared with the information available  on nebulae around LBVs, the nebula 
around R4 appears more complicated in nature. From the kinematics, it is fair 
to conclude that:
\begin{itemize}
\item the nebula is  {\it not} a simple  expanding shell.  An expanding shell 
would result in a radial velocity map which has maximum dispersion in 
correspondence to the position of the star, and minimum velocity at the
shell boundaries (eg. AG Car: Smith 1991, Nota et al. 1992). 
\item the nebula is not strictly bipolar.  Compared with the radial velocity 
maps derived for HR Car, a prototypical bipolar outflow (Nota et al. 1997), 
the situation is very different: in the case of HR Car, there is a clear 
demarcation between  the two sides of the bipolar outflow, with the redshifted 
region limited to the NW quadrant, and the blueshifted  to the SE quadrant 
(see Figure 9 in Nota et al. 1997).
\end{itemize}
In order to provide a consistent explanation for the radial velocity maxima 
and minima observed in the R4, a more complicated structure needs to be 
invoked, 
which is symmetrical around two axes. In the bottom panel of Figure 3, 
we provide 
a cartoon of one possible structure, in which a cloverleaf morphology
is aligned with the peculiar radial velocity features. In this proposed  
structure, the outflow occurs in four directions along two axes, perpendicular 
to each other, and oriented  at a PA $\simeq$ 45$^{o}$.  This complicated 
structure, although speculative, has been observed in planetary nebulae.
Only a direct image will confirm  whether such speculation is correct. 

\subsection{Nebular composition}
In addition to the kinematical properties, the nebular lines provide
also some information on the chemical composition of the nebula.  We have 
computed the
line ratio [NII]6584/[SII]6717 for the R4 nebula and the local
interstellar medium from the long-exposure spectra which we corrected for
atmospheric extinction. We
derived a  ratio [NII]6584/[SII]6717 of about 3 (in
agreement also with the data of Zickgraf et al. 1996) and 0.3 for the R4
nebula and the local interstellar medium, respectively. Such result
 is inconsistent with the same line ratio measured in HII regions and SN
remnants in the SMC by Russell \& Dopita (1990). Typically,
the [NII]6584/[SII]6717 ratio is 0.6 for the HII regions
(with one exception: N84C is characterized by a value of 2.4) and varies
between 0.2 and 0.4 in the case of SNR, independently of the position in
the galaxy. A 
[NII]6584/[SII]6717 value of 0.3 is considered to reflect the intrinsic lower 
N content of the SMC with respect to the Galaxy and the LMC. We may then 
conclude that the R4 nebula is N-enriched by a factor 10 with respect to both 
the local interstellar medium and HII regions/SN remnants.

\section{Discussion}
>From Section~3, we conclude that R4 is surrounded by  a bipolar circumstellar 
nebula, nitrogen enriched, with a dynamical age of $\sim$ 1.2 $\times$ 
10$^4$yr.
This nebula appears to have been ejected from the central star, and its
morphological, kinematic and chemical properties are comparable to
the average properties of LBV nebulae (cf., Nota et al. 1995).
Although Esteban \& Fernandez (1998) have detected a circumstellar nebula 
around the galactic B[e] star MCW$\,$137 which appears not to be 
chemically enriched, our findings for  R4 provide
the first evidence for an ejected nebula around a B[e]~supergiant.

\subsection{The progenitor evolution of R4}

In order to investigate the implications of our finding for the connection
of B[e]~supergiants and LBVs, let us recall the proposed evolutionary scenarios
for the B[e]~supergiants. Table~3 summarises the expected stellar and
circumstellar nebula properties of B[e]~supergiants according to the 
very massive main sequence star scenario, the blue loop scenario
and the binary merger scenario (Langer \& Heger 1998).

In the first scenario, the star reaches the $\Omega$-limit (cf. Section~1)
during its
main sequence evolution due to its high luminosity, i.e. its proximity to 
the Eddington limit (Langer 1998). This appears possible only for the
most massive stars. Since at low metallicities the Eddington-limit  
is even higher 
at higher luminosity than for young stars in the solar neighborhood 
(Ulmer \& Fitzpatrik 1998), this scenario would require an extraordinarily 
fast rotation to apply to the case of R4.

In the second scenario, the star reaches the $\Omega$-limit on a blue loop
evolving off the Hayashi line during core helium burning (Heger \& Langer
1998). This scenario can not apply to R4 since stars on blue loops do not
exceed effective temperatures of $\sim 20\, 000\,$K (cf. 
Langer \& Maeder 1995), i.e., the blue loops never extend to the main sequence
band.

In the third scenario, an equatorial disk or ring is created by mass overflow 
through the second Lagrangian point in a close binary system in the course
of a close binary merger. This scenario 
was supposed to be most appropriate for the
case of R4 by Langer \& Heger (1998) for the following reasons.
The B[e]~supergiant R4 has an evolved A~type companion star with a mass of 
about 12.9$\mso$ (Zickgraf et al. 1996). While the B[e]~star mass has been
determined to a similar value ($\sim 13.2\mso$) its bolometric luminosity
outshines that of the A~star by a factor of $\sim 10$.
Zickgraf et al. conclude from the high luminosity of R4 ($\sim 10^5 \lso$),
and from its strong surface enrichment in CNO processed material,
that it must have lost large amounts of mass in a previous red 	supergiant 
phase.

However, as noted by Langer \& Heger (1998), even if a $\sim 20\mso$ red
supergiant at the very low metallicity of the SMC had lost 
about $10\mso$ via its stellar wind (which appears unlikely 
for several reasons; e.g., the large envelope mass of the progenitor of
supernova 1987A), there would remain a basic puzzle in the R4 binary system.
The A~star is clearly beyond core hydrogen exhaustion: how then
can the B~star have an {\em evolved} companion which is ten times less 
luminous?
The B~star should have long exhausted its fuel and exploded as a supernova.
Clearly, binary mass transfer must have occured in this system.
Now, the orbital separation in the system is presently about 23~A.U.
or 5000$\rso$ (Zickgraf et al. 1996), which may be too large to allow any
mass transfer from the A~star to the B~star. Also, mass transfer in very wide
systems is supposed to be unstable and would not leave the two stars at a
large separation (e.g., Podsiadlowski et al. 1992), and can therefore
be excluded.

A viable binary scenario for R4 may be that the B[e]~star
has been formed by a recent binary merger, and the A~star is not involved in
any mass transfer but only serves as a suitable clock (Langer \& Heger 1998).
Wellstein et al. (2000) find in a parameter study of binary evolution
models that a system which starts out with a 12$\mso$ and a 11$\mso$
star on a 40 day orbit would evolve into contact after core hydrogen exhaustion
in the 12$\mso$ star, which leads to mass overflow through the outer Lagrangian
point $L_2$ and then most likely to a merger. The $L_2$ overflow, which is
likely to comprise several solar masses of material (Wellstein et al. 2000), 
gives rise to a circumstellar nebula, which is then
shaped by the ensueing B~star wind,
like the Homunculus nebula around $\eta$~Carinae in the wind interaction
scenario of Langer et al. (1999).
The large amount of angular momentum
in the stellar merger remnant may lead to a disk wind according to the 
Bjorkman-Cassinelli mechanism, which then is responsible for the
B[e]~morphology of the stellar spectrum. Furthermore, the merger star would be
expected to contain an overly large helium fraction in its interior
--- which would give it an unusually large L/M-ratio ---
and a surface strongly enriched in CNO products.

That all these details are observational facts makes R4 the strongest
massive star candidate for a binary merger. Within this picture, the system
started out as a triple system with three very similar stars, a close
pair of, say, 12$\mso$ and 11$\mso$, in a wide orbit with a $\sim 13\mso$
star. Soon after the latter has evolved off the main sequence into an
A~type supergiant, the close pair would merge to form the B[e]~supergiant,
about 1.2 $\times$ 10$^4$ yr ago. 

\subsection{Implications for other B[e]~supergiants and LBVs}

In order to understand the relevance of R4 for massive stars in general,
we should establish whether R4 is a peculiar or a typical B[e]~supergiant.
The latter case would have the dramatic implication that most B[e]~supergiants
might be the result of a binary merger. However, there are at least two
arguments against this proposition. First,
R4 has an extreme location in the HR diagram compared to all other 
B[e]~supergiants in that its location is by far the closest to the main
sequence band. Zickgraf et al. (1996) compared it with evolutionary tracks of
Charbonnel et al. (1993), where R4 falls exactly on the terminal age main
sequence of a 20$\mso$ track. Second, R4 is today the only B[e]~supergiant
with an ejected circumstellar nebula. Before generalizing the evolutionary 
scenario of its progenitor, one would certainly want to
detect more examples with ejected nebulae. 

On the other hand, the similarity of the properties of the R4 nebula and that
of LBV nebulae in general --- which provides a remarkably homogeneous class
--- is striking.  These nebulae  are all  very similar
in terms of morphological and  physical properties. They  are  all typically
1 parsec in size, with morphologies which are mildly to extremely
bipolar.  They expand in the
surrounding medium with velocities of the order of 50 -- 100 km
s$^{-1}$.  Their size and expansion velocities identify dynamical
timescales which are of the order of several thousand of years. 
Densities   are generally found to be low
(500 - 1000 cm$^{-3}$) and so are temperatures, found in the range 5000
- 10000 K. In terms of overall physical
and chemical properties, the nebula surrounding R4 would fit this category 
well.
However,
two facts preclude the possibility that the R4 nebula
and all LBV nebulae have been formed the same way.
The first is R4's location in the HR~diagram, which of all the 
stars in the group of
luminous B[e]~supergiants is farthest away from the LBV regime.
The second, even stronger argument is the fact that some 
LBV nebulae appear to have multiple
shells (Nota et al. 2000) and thus the central stars
most likely experienced multiple outbursts,
which appears not possible within the binary merger scenario.


Although we argue that not all B[e] supergiants are formed by stellar mergers,
we conclude that at least two intrinsically very different
nebula formation mechanisms, the binary merger and the single star
LBV outburst mechanism, can produce nebulae with very similar properties.
Seemingly, the prerequisites for the nebular structure are the same in both
cases, i.e. a massive disk ejected by the central star which is then shaped
by its strong wind. While the disk forms through the $L_2$-Roche lobe
overflow in the case of the merger (hardcore hydrodynamic calculations
for such events exist so far only for low mass stars, which, however,
support the general idea; cf. Yorke et al. 1995),
it may be formed through rotational wind compression (Bjorkman \& Cassinelli
1993) in the single star case (cf. Langer et al. 1999).

\section{Conclusions}
Our longslit spectroscopic observations reveal that R4 is embedded in a
circumstellar nebula whose full spatial extension is 8.6$''$, i.e. 
2.4 pc at the SMC distance of 57.5 Kpc (van den
Bergh 1989). The emission lines show a significant structure in 
velocity indicating that the nebula is expanding at 100 km
s$^{-1}$ on average. This implies a dynamical age of 1.2 $\times$ 10$^4$ yr.
The radial velocity maps obtained for the nebula at 3$''$ North and
3$''$ South from the central star are characterized by two velocity maxima  
and two velocity minima, respectively, located at the same positions
with respect to the star. Hence, the two velocity distribution
appear to be symmetric not only in the velocity field but also along the
E-W direction (cf. Figure 3). Such a symmetry 
excludes that the R4 nebula is either a simple expanding shell
(i.e. AG Car) or a simple bipolar outflow (i.e. HR Car). It rather
suggests a cloverleaf morphology, where the outflow occurs in four
directions along two axes, perpendicular to each other and oriented at
PA $\simeq$ 45$^o$. 
\par\noindent
Since the line flux ratio [NII]6584/[SII]6717 is 
almost independent of electron temperature and density, it can be used
to estimate any overabundance of nitrogen with respect to the
"unprocessed" sulphur. The nebular [NII]6584/[SII]6717 ratio turns out
to be 3 against a value of 0.3 as measured for the local interstellar
medium in the same observed spectra, and a value between 0.6 and 0.2 as
derived for HII regions and SN remnants in the SMC by Russell \& Dopita
(1990). This factor of 10 discrepancy clearly indicates that the R4
nebula is nitrogen enriched and therefore, since it is also kinematically
associated with the central B[e] supergiant, it is an ejected
nebula. {\it R4 is surrounded by a bipolar and N-enriched nebula,
ejected from the central star, whose morphological, kinematical and
chemical properties well compare with the mean properties of LBV
nebulae.} 
\par\noindent
We have shown that the central
star most likely formed through a binary merger, as proposed by Langer \& 
Heger (1998). This makes R4 the strongest observational counterpart of such 
event among massive stars, which has long been sought for. E.g., in a 
comprehensive binary and single star evolution and population sysnthesis study,
Podsiadlowski et al. (1992) conclude that $\sim 25$\% of all massive binaries
undergo a merging process, most likely so just after the initially more 
massive star has terminated core hydrogen burning. 
\par\noindent
In Section~4.2, we concluded from the properties of R4 that two
distinct mechanisms can form circumstellar nebulae of exactly that type
found around LBVs. This has two consequences. First, it implies that
some LBV outbursts and nebulae may in fact be due to the merging process
of two massive stars. The predicted amount of expected binary mergers 
(see above) implies that this is in fact likely. Second, it may
deepen the understanding of why bipolar circumstellar nebulae
are such a frequent phenomenon. E.g., what holds for massive stars may be
true for the progenitors of planetary nebulae, and in the end the dispute
of whether bipolar planetaries are formed through binaries (e.g., Soker 
1998) or single stars (e.g., Garc\'{\i}a-Segura et al. 1999) may end in
a draw. 
\par
We conclude by urging for imaging observations of the R4 nebula. They will
not only provide for the first time the morphological details of a nebula
ejected by a massive stellar merger and will thereby allow to constrain
the hydrodynamical processes at work in such phenomenon, but it will
perhaps reveal clues of how to empirically discriminate binary and single star
ejection mechanisms and thus allow for a better understanding of bipolar 
circumstellar nebulae in general.

\acknowledgments
The authors wish to thank Guillermo Garc\'ia-Segura and the staff of the
Tonantzintla Institute for their hospitality during their visit where
this project was perceived. AP, AN and MC would like to thank STScI and 
ST-ECF for their hospitality during their visits where the paper was
finalized. NL acknowledges supporting from    
the Deutsche Forschungsgemeinschaft through grants La~587/15-1 and 16-1.
RSL acknowledges funding from the HST grant to GO proposal 6540.

\clearpage
\begin{deluxetable}{c c c c c c c c}
\tablecaption{Journal of the observations.}
\tablewidth{0pt}
\tablehead{
\colhead{Date} & \colhead{Grating} & \colhead{$\lambda$ range} &
\colhead{Dispersion} & \colhead{Resolution} & \colhead{Exp. Time}
& \colhead{Location}  \nl
\colhead{(1998)} & \colhead{} & \colhead{(in \AA)} & \colhead{(in
\AA/pix)} & \colhead{(FWHM in \AA)} & \colhead{(in s)} & \colhead{}
}
\startdata
July 30 & 7 & 4470 - 5820 & 0.66 & 2.72 & 1800 & on the star \nl
July 31 & 7 & 5080 - 6420 & 0.66 & 2.36 & 1500 & 3$''$ North \nl
   "    & 6 & 6240 - 6870 & 0.31 & 1.12 & 60& 3$''$ North \nl
   "    & " &      "      &  "   &  "   & 1800 & "  \nl
   "    & " &      "      &  "   &  "   & 80 & on the star \nl
   "    & " &      "      &  "   &  "   & 60 & 3$''$ South \nl
   "    & " &      "      &  "   &  "   & 1800 & "  \nl
\enddata
\end{deluxetable}

\clearpage
\begin{deluxetable}{c c c c}
\tablecaption{Comparison of observable properties predicted by the three
B[e]~supergiant evolutionary scenarios discussed in this paper 
(cf. Langer \& Heger 1998).}
\tablewidth{0pt}
\tablehead{
\colhead{} & \colhead{very massive main} &\colhead{supergiant on} 
&\colhead{single star} \nl
\colhead{} & \colhead{sequence star at} &\colhead{blueward excursion} 
&\colhead{remnant of} \nl
colhead{} & \colhead{the $\Omega$-limit} &\colhead{from Hayashi line} 
&\colhead{binary merger}} 
\startdata
            &                      &                & 10$\,$000$\,$K ... \nl
$\teff$     & $\simgr 20\, 000\,$K &$\simle 20\, 000\,$K& 30$\,$000$\,$K \nl
   ~        &      ~               &       ~        &         ~           \nl
luminosity  & $\simgr 10^5 \lso$    &$\simle 10^5 \lso$&$\simgr 10^4 \lso$ \nl
   ~        &      ~               &       ~        &         ~           \nl
time scale  & some 10$^5$yr        & some 10$^4$yr  &      (?)            \nl
   ~        &      ~               &       ~        &         ~           \nl
time integr.&  $\sim 5\mso$        & $\sim 0.1\mso$ &  $\sim 5\mso$(?)    \nl
disk mass   &      ~               &       ~        &         ~           \nl
            &      ~               &       ~        &         ~           \nl
nebular chem.& non-enriched to     & moderately     & strongly            \nl
composition &moderately enriched   & enriched       & enriched            \nl
            &      ~               &       ~        &         ~           \nl
\enddata
\end{deluxetable}

\clearpage
\begin{figure}
\plotone{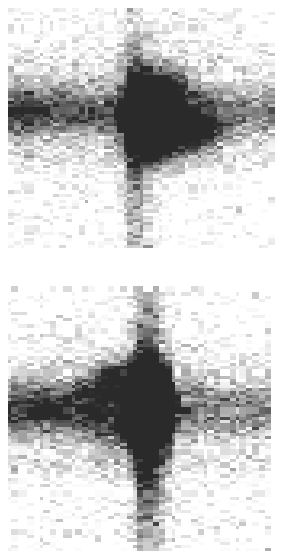}
\figcaption{Top panel: the [NII]$\lambda$6584 line as it
appears in the longslit spectrum taken at 3$''$ North. Wavelength
is on the abscissae axis, increasing from left to right, while
the spatial direction is on the ordinates axis. Bottom panel:
the [NII]$\lambda$6584 line in the longslit spectrum taken at 3$''$ South.}
\end{figure}

\clearpage
\begin{figure}
\plotone{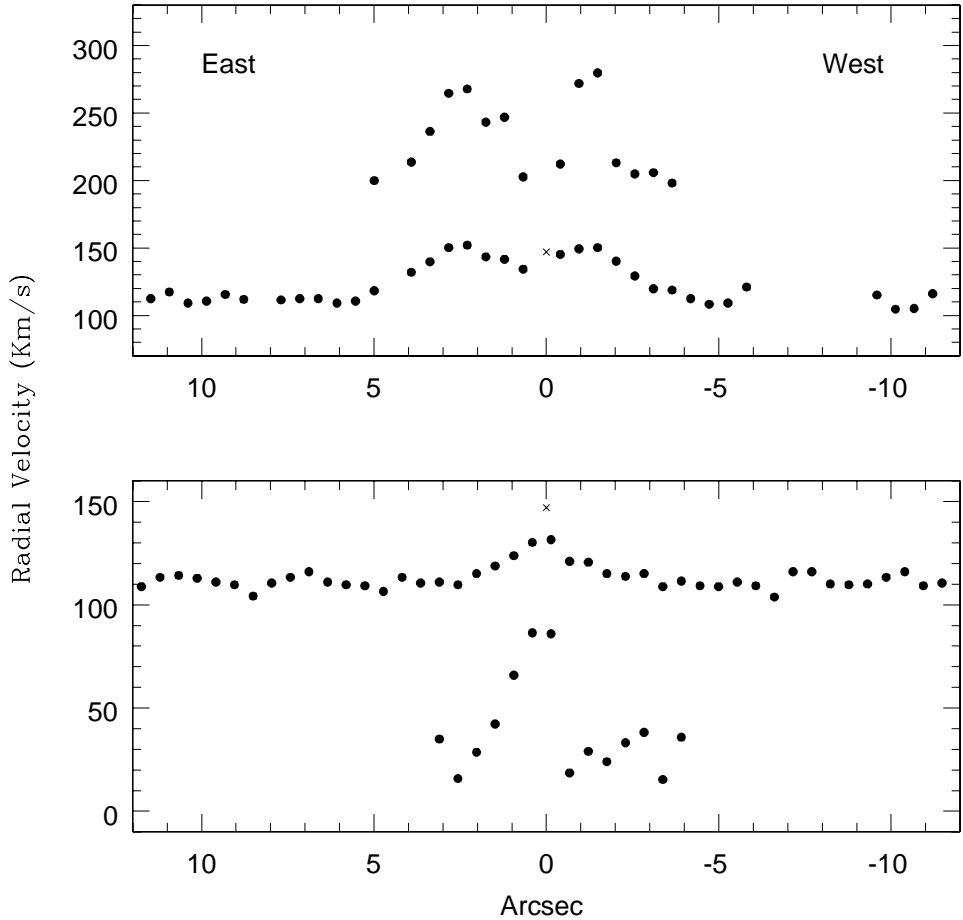}
\figcaption{Top panel: radial velocities (in the
heliocentric system) measured from the [NII]$\lambda$6584 as a function
of position with respect to the star (at 0), in the longslit spectrum at
3$''$ North. Bottom panel: radial velocities measured from the same line
in the longslit spectrum obtained at 3$''$ South. The cross represents
the star velocity of 147 km s$^{-1}$ as derived by Zickgraf et al.
(1996).}
\end{figure}

\clearpage
\begin{figure}
\figcaption{A cartoon of the R4 nebula. The speculated structure
is of a cloverleaf, where the outflow occurs in four directions along two axis,
perpendicular to each other and oriented at PA $\simeq$ 45$^o$ (bottom
panel). This structure is aligned with the two velocity maxima in
the spatial distribution of radial velocities obtained at 3$''$ North
and with the two velocity minima of the radial velocity map measured at
3$''$ South (top panel).}
\end{figure}
\end{document}